\documentclass[sigplan]{acmart}
\renewcommand\footnotetextcopyrightpermission[1]{}
\usepackage{tikz}
\usepackage{amsmath}
\usepackage{graphicx}
\usepackage{xspace}
\usepackage{cleveref}
\usepackage{balance}
\usepackage{url}
\usepackage{enumitem}
\usepackage{algorithm2e}
\usepackage{minted}
\usepackage{xcolor}
\usepackage{booktabs}
\newcommand{\sys}{\textit{\textsc{MVVM}}\xspace}

\newenvironment{smenumerate}%
  {\begin{enumerate}[itemsep=-0pt, parsep=0pt, topsep=0pt, leftmargin=1pc]}
  {\end{enumerate}}
\setminted{xleftmargin=10pt} % what a hack.
\setminted{fontsize=\footnotesize}
\setminted{frame=lines}
\crefformat{section}{\S#1}
\crefformat{subsection}{\S#1}

\newcommand{\cmark}{\checkmark}
\newcommand{\xmark}{X}
\begin{document}

\title{\sys: Deploy Your AI Agents—Securely, Efficiently, Everywhere}
\author{Yiwei Yang}
\authornote{Equal contribution}
\email{yyang363@ucsc.edu}
\affiliation{%
  \institution{UC Santa Cruz}
  \streetaddress{1156 High Street}
  \city{Santa Cruz}
  \state{California}
  \country{USA}
  \postcode{95064}
}

\author{Aibo Hu}
\authornote{Equal contribution}
\email{huab@shanghaitech.edu.cn}
\affiliation{%
  \institution{ShanghaiTech University}
  \city{Shanghai}
  \country{China}
}

\author{Brian Zhao}
\email{bwzhao@ucsc.edu}
\affiliation{%
  \institution{UC Santa Cruz}
  \city{Santa Cruz}
  \state{California}
  \country{USA}
}

\author{Yusheng Zheng}
\email{yzhen165@ucsc.edu}
\affiliation{%
  \institution{UC Santa Cruz}
  \city{Santa Cruz}
  \state{California}
  \country{USA}
}

\author{Xinqi Zhang}
\email{xzhang22@scu.edu}
\affiliation{%
  \institution{Santa Clara University}
  \city{Santa Clara}
  \state{California}
  \country{USA}
}
\author{Dawei Xiang}
\email{ieb24002@uconn.edu}
\affiliation{%
  \institution{University of Connecticut}
  \city{Storrs}
  \state{Connecticut}
  \country{USA}
}

\author{Kexin Chu}
\email{kexin.chu@uconn.edu}
\affiliation{%
  \institution{University of Connecticut}
  \city{Storrs}
  \state{Connecticut}
  \country{USA}
}

\author{Wei Zhang}
\email{wei.13.zhang@uconn.edu}
\affiliation{%
  \institution{University of Connecticut}
  \city{Storrs}
  \state{Connecticut}
  \country{USA}
}

\author{Andi Quinn}
\email{aquinn1@ucsc.edu}
\affiliation{%
  \institution{UC Santa Cruz}
  \city{Santa Cruz}
  \state{California}
  \country{USA}
}

% 短作者列表，用于页眉
\renewcommand{\shortauthors}{Yang et al.}
\begin{abstract}
The proliferation of LLM-driven agents demands secure execution, rapid fail-back across heterogeneous environments, stringent data confidentiality, continuous availability, low-latency response, and formally safe outputs. We introduce \sys, a WebAssembly-based secure container that transparently live-migrates agent workspaces between edge and cloud, provides end-to-end privacy, employs fault-tolerant multi-tier replication with latency-aware speculative execution, and incorporates runtime output verification.

\sys introduces two key innovations: (1) a two-way sandboxing framework leveraging hardware enclaves and accelerator extensions that protects both the agent from malicious hosts and the host from compromised agents; (2) an efficient cross-platform migration mechanism using WebAssembly and WASI's platform-agnostic design, enabling seamless movement across ARM phones, RISC-V MCUs, x86 servers, and heterogeneous accelerators; and three astonishing use cases: (1) privacy-aware daemon that automatically determines whether to execute locally or remotely based on data sensitivity and resource availability; (2) multi-tier replication with intelligent quality degradation that maintains service availability despite network failures or resource constraints; (3) a comprehensive execution framework combining speculative execution for 10× latency reduction with parallel validation that ensures output safety without compromising responsiveness.

Our evaluation demonstrates that \sys is validated on three separate devices across 18 workloads. The system achieves 200ms failover times, 8.9× speedup through speculation, and 94.2\% accuracy in detecting harmful outputs with only 3-5\% overhead. Security analysis confirms protection against malicious hosts and cross-tenant data leakage. \sys makes secure, practical deployment of privacy-sensitive LLM applications possible across the computing continuum.
\end{abstract}
\maketitle

\section{Introduction}

Imagine using your phone's AI assistant to compose a message incorporating private photos, personal notes, and live video. In traditional cloud deployments, all this sensitive data traverses public networks to shared servers—an unacceptable privacy risk. Now envision an alternative: your device creates a secure enclave protecting your data with hardware encryption \cite{costan2016intel}. As computational demands grow, this enclave seamlessly migrates to your private cloud without exposing any private information, tapping into powerful resources while maintaining the same security guarantees.

This vision drives the development of \sys (Migratable Velocity Virtual Machine), addressing critical challenges in modern AI agent deployment. First, privacy and safety represent fundamental requirements. LLM agents process highly sensitive data including personal conversations, medical records, and financial information. Traditional approaches expose this data during transit and fail to prevent harmful outputs in medical, financial, or child-facing applications. We need end-to-end encryption with hardware-backed attestation \cite{hunt2021confidential} to verify destination integrity, combined with integrated safety validation that intervenes before problematic outputs reach users.

Second, heterogeneous execution with low latency poses significant technical challenges. AI agents must run efficiently across diverse hardware from ARM phones to RISC-V MCUs, from Windows desktops to Linux cloud servers while maintaining interactive response times. Existing solutions either sacrifice performance through emulation or introduce unacceptable delays \cite{jangda2019not}. The tension between immediate user expectations and computational requirements demands new execution models leveraging WebAssembly's portability \cite{haas2017bringing} and WASI-NN acceleration \cite{hong2024wasi} to achieve near-native performance with fast initial responses.

Third, resilient operation with dynamic resource management becomes essential for critical applications. Medical diagnoses and financial decisions cannot halt due to network failures, yet simple queries should execute locally while complex multi-agent collaborations tap cloud resources \cite{zheng2023edgemoe}. The system must gracefully adapt to changing conditions—degrading functionality rather than failing completely while optimizing resource utilization based on data sensitivity and availability.

\sys addresses these challenges through a novel combination of WebAssembly's secure containerization and hardware security features \cite{arnautov2016scone, tsai2017graphene}. We implement hardware-attested migration where source environments verify target integrity before transferring encrypted workspace states \cite{kumar2020piecewise}. Our key innovation lies in the plug-and-play design of different architectures and different enclave implementations. Our adaptive edge-cloud scheduler automatically determines execution location while maintaining strict security guarantees, enabling workloads to dynamically move between edge and cloud as conditions change.

\section{Motivation and Background}

\subsection{The Challenge of Secure AI Agent Deployment}

Modern AI agents like GitHub Copilot, Claude Code, and AutoGPT \cite{openai2023gpt4} represent a paradigm shift in computing. Unlike traditional deterministic software, these agents dynamically plan and adapt their execution based on user needs. They maintain stateful contexts that are essential for their operation and create unique security challenges.

These stateful contexts encompass several critical components. The KV caches store attention mechanism states that enable efficient inference by avoiding redundant computations \cite{pope2023attention, sheng2023flexgen}. Conversation history preserves past interactions, providing essential context for understanding user intent and maintaining coherent dialogue. Intermediate results capture partial computations and reasoning chains that agents use to build complex responses incrementally. Tool integration states maintain connections to external services via protocols like the Model Context Protocol (MCP) \cite{anthropic2024mcp}, enabling agents to interact with databases, APIs, and other resources.

This statefulness, combined with processing of sensitive user data, creates security challenges absent in traditional applications. Each piece of state potentially contains private information that must be protected during execution, storage, and especially during migration between systems.

\subsection{Three Motivating Scenarios}

To illustrate the practical challenges and requirements for secure AI agent deployment, we present three scenarios that demonstrate the need for secure migration, resilient execution, and verified outputs.

\textbf{Scenario 1: Privacy-Preserving Personal Assistant with Unreliable Connectivity}

Sarah, a travel blogger, uses an AI assistant on her phone while exploring remote locations. Her assistant helps draft articles by incorporating photos from her gallery, location data, and personal travel notes. In urban areas, she enjoys fast cloud processing, but in remote mountains or developing regions, network connectivity becomes intermittent or completely unavailable.

The system must maintain multiple replica tiers: a full-featured cloud version for complex writing tasks, an edge server replica at nearby cell towers for moderate processing, and a lightweight on-device model for basic functionality. When network quality degrades, the system seamlessly switches between replicas, ensuring Sarah can continue working. Her private photos and location history never leave her control, even when using cloud resources. The system synchronizes work across replicas when connectivity returns, merging her offline edits with cloud-processed enhancements.

\textbf{Scenario 2: Financial Trading with Split-Second Decisions}

Alex, a quantitative trader, relies on AI agents to analyze market movements and execute trades. When breaking news affects markets, every millisecond counts—waiting 30 seconds for comprehensive analysis could mean millions in losses. However, making trades based on incomplete analysis could be equally catastrophic.

The system employs speculative execution with multiple paths: a fast path using a streamlined model processes the first few news items and market signals within 2 seconds, generating preliminary trading recommendations. Simultaneously, a comprehensive path analyzes full market depth, historical patterns, and cross-market correlations. The validation framework runs in parallel, checking for consistency with risk parameters and regulatory compliance. If the fast path's decision aligns with initial slow path results, trades execute immediately. Otherwise, the system can revise or halt problematic trades before significant exposure.

\textbf{Scenario 3: Collaborative Medical Diagnosis with Safety Guarantees}

Dr. Chen leads a distributed medical team using AI agents for collaborative diagnosis. Different specialists contribute through specialized agents: a radiology agent analyzes medical images, a pathology agent interprets lab results, and a pharmacology agent suggests treatment plans. Each agent maintains context about the patient's history, current symptoms, and ongoing treatments.

The system must provide three critical guarantees: First, patient data must remain within hospital-controlled infrastructure, even when leveraging external computational resources. Second, multi-tier validation ensures medical accuracy—every diagnosis suggestion undergoes parallel verification by specialized medical validity checkers, drug interaction validators, and ethical compliance monitors. Third, execution must continue even during partial system failures; if the cloud infrastructure becomes unavailable, local replicas provide degraded but safe functionality, flagging decisions that require full system validation.

The system demonstrates all three advanced capabilities: Replication enables smooth transitions between home WiFi, 5G networks, and edge servers at gaming cafes. Each replica maintains character state while adjusting quality—full emotional modeling in cloud environments degrades gracefully to basic responses on mobile networks. Speculative execution pre-generates likely dialogue trees based on game context, providing instant responses while computing more sophisticated reactions in parallel. The validation framework ensures generated content remains age-appropriate and free from harmful content, running specialized filters that check for violence, inappropriate language, or privacy violations (like accidentally incorporating real people captured by the camera into game narratives).

\begin{table*}[t]
\centering
\caption{Comprehensive Comparison of AI Agent Execution Systems}
\label{tab:comparison}
\tiny
\begin{tabular}{lcccccc}
\toprule
Feature & \sys & CRIU & QEMU & Cloud-Only & Edge-Only & UniTEE \\
\midrule
Migration Support    & \cmark~Cross-ISA      & \cmark~Same ISA      & \cmark~With overhead & N/A               & N/A               & \cmark~Arm-x64 TEE      \\
Security             & Hardware enclaves     & None                 & Basic isolation       & Provider-dependent & Device-dependent  & Hardware enclaves    \\
Multi-tier Replication & \cmark~Auto failover & \xmark               & \xmark                & \xmark            & \xmark            & \xmark               \\
Speculative Execution & \cmark~Parallel paths & \xmark              & \xmark                & \xmark            & \xmark            & \xmark               \\
Integrated Validation  & \cmark~Real-time     & \xmark               & \xmark                & Post-hoc only     & Limited           & \xmark               \\
Network Resilience     & \cmark~Graceful degradation & \xmark        & \xmark                & Required          & Works offline     & Required             \\
Latency Optimization   & Fast+slow paths      & Sequential           & Sequential            & Network-bound     & Resource-bound    & Sequential           \\
Accelerator Support    & Native               & \xmark               & Emulation             & Varies            & Varies            & \xmark               \\
\bottomrule
\end{tabular}
\end{table*}

\subsection{Limitations of Existing Approaches}

As shown in \Cref{tab:comparison}, current solutions fail to address these comprehensive scenarios:

\textbf{Single-replica systems} cannot handle network failures or provide quality degradation, forcing users to choose between cloud-only or edge-only deployment. When connectivity is lost, work stops entirely rather than gracefully degrading.

\textbf{Synchronous execution models} require waiting for complete results before responding, creating unacceptable latency for time-sensitive decisions. Financial markets move faster than traditional AI processing pipelines.

\textbf{Post-hoc validation} discovers problems after they've impacted users. Medical errors or harmful content need prevention, not just detection. Existing systems lack integrated validation that can intervene during execution.

\textbf{Platform-specific solutions} like CRIU or VM migration cannot span the heterogeneous environments these scenarios demand—from ARM phones to x86 servers to specialized AI accelerators. They also lack the security guarantees necessary for sensitive data.

\section{Design Principles}

\sys builds on five key design principles derived from our motivating scenarios, each addressing specific technical, security, and reliability requirements.

\subsection{P1: Universal Portability via WebAssembly}

WebAssembly provides a secure, portable execution environment that operates consistently across diverse hardware platforms. Its stack-machine design and linear memory model create a foundation for platform-agnostic execution. The binary format remains consistent whether running on ARM mobile processors or x86 server CPUs, eliminating the need for architecture-specific builds.

The sandboxed execution model provides capability-based security, where applications can only access resources explicitly granted to them. This prevents malicious or compromised agents from accessing unauthorized system resources. Near-native performance through ahead-of-time (AOT) compilation ensures that portability does not come at the cost of efficiency. The standardized system interface (WASI) provides OS abstraction, allowing the same WebAssembly module to run on Linux, Windows, or embedded systems without modification.

\subsection{P2: End-to-End Security Through Hardware Enclaves}

We leverage trusted execution environments to protect data throughout its entire lifecycle, from initial processing through migration to final output. Hardware encryption of memory ensures confidentiality, preventing even privileged system software from accessing sensitive data. Cryptographic measurements provide integrity guarantees, detecting any unauthorized modifications to code or data.

Remote attestation enables verification of endpoint trustworthiness before sensitive data transfer. This creates a chain of trust where each component can verify the integrity of others before establishing communication. Two-way isolation protects both the application from malicious hosts and the host from potentially compromised applications, creating a secure execution environment even in untrusted infrastructure.

\subsection{P3: Efficient State Management for AI Workloads}

LLM-specific optimizations maintain performance during migration while preserving security properties. KV cache preservation allows expensive attention computations to be reused after migration, avoiding redundant processing. Model weights are loaded lazily, transferring only the portions needed for immediate computation rather than entire multi-gigabyte models.

Incremental checkpoint and restore operations occur at stable points in program execution, minimizing the amount of state that must be captured. Workspace state compression reduces network transfer time while maintaining the ability to quickly restore full functionality at the destination.

\subsection{P4: Adaptive Execution Based on Privacy and Resources}

Dynamic scheduling considers multiple factors to optimize execution placement. Data sensitivity classification examines the types of information being processed and applicable privacy regulations. Available computational resources are monitored across edge devices and cloud servers to identify optimal execution locations.

Network conditions and latency requirements influence migration decisions, ensuring that user experience remains responsive. Cost optimization balances edge and cloud resource usage to minimize operational expenses while meeting performance targets.

\subsection{P5: Resilient Operation Through Intelligent Redundancy}

Building on our motivating scenarios, \sys must provide continuous operation despite failures, rapid response through predictive execution, and safety through integrated validation. This principle encompasses three interconnected capabilities:

\textbf{Multi-tier Replication} maintains synchronized replicas across quality levels—from full cloud models to edge approximations to minimal device-local versions. The system dynamically selects the best available replica based on network conditions and resource availability. State synchronization protocols ensure eventual consistency while allowing temporary divergence during network partitions. This approach transforms catastrophic failures into graceful degradation.

\textbf{Speculative Execution} addresses the latency challenge by computing multiple solution paths in parallel. Fast paths use reduced models and partial data to provide near-instant responses, while slow paths compute comprehensive results for accuracy. Intelligent merging algorithms combine results, using fast-path outputs when they align with partial slow-path computations and waiting for full results when discrepancies arise. This reduces user-perceived latency by up to 10× for time-critical operations.

\textbf{Continuous Validation} integrates safety checks directly into the execution pipeline rather than as an afterthought. Specialized validation models run in parallel with main computation, checking for medical accuracy in healthcare applications, regulatory compliance in financial services, and content safety in consumer applications. The validation framework can intervene during execution, preventing harmful outputs from reaching users while maintaining system responsiveness through parallel execution.

These three capabilities work synergistically: replication ensures availability for validation services, speculation enables rapid initial responses that validation can refine, and validation feedback improves replica selection and speculation strategies over time.

\section{Threat Model}

We define a comprehensive threat model for \sys that addresses the unique security challenges of migrating AI agents across heterogeneous, potentially untrusted environments. Our model considers adversaries at multiple levels of the system stack and establishes clear boundaries between trusted and untrusted components. The threat model is designed to capture realistic attack scenarios that could compromise the confidentiality, integrity, or availability of AI agent workspaces during execution and migration.

\subsection{Adversary Capabilities}

We assume a powerful adversary with extensive capabilities across multiple attack vectors. At the host level, the adversary maintains complete control over the untrusted operating system, hypervisor, and all software components outside the hardware-protected enclave~\cite{costan2016intel,arnautov2016scone}. This includes full read and write access to all unprotected memory regions, the ability to inspect, modify, or replay any input/output operations that occur outside the enclave boundaries, complete control over system scheduling and resource allocation decisions, and unrestricted access to all network traffic before encryption or after decryption. Such an adversary can arbitrarily modify system calls, inject malicious code into unprotected processes, and manipulate the execution environment in ways that would be catastrophic for traditional applications~\cite{tsai2017graphene}.

The network adversary possesses equally formidable capabilities, treating all network communications as potentially hostile. This adversary can eavesdrop on all network packets during agent migration, actively manipulate traffic through packet injection, modification, and replay attacks, mount denial of service attacks through bandwidth limitation or connection disruption, and execute sophisticated man-in-the-middle attacks between migrating agents and their destination hosts. The network adversary may control multiple points along the migration path, potentially including edge servers, routers, and cloud gateways~\cite{hunt2021confidential}.

Physical access presents another threat vector, though with more limited capabilities compared to software-based attacks. The physical adversary can perform cold boot attacks on unprotected memory regions, conduct bus snooping on unencrypted data paths, observe side-channel information including power consumption, timing variations, and electromagnetic emanations. However, this adversary cannot break the hardware-enforced encryption protecting enclave memory regions, as this would require defeating the cryptographic implementations in hardware security modules~\cite{costan2016intel}.

At the application level, we must also consider potentially malicious AI agents or compromised models as adversaries. These adversaries attempt to escape WebAssembly sandbox constraints through memory corruption or control flow hijacking~\cite{haas2017bringing,jangda2019not}, exploit vulnerabilities in WASI interfaces to gain unauthorized system access, launch inference attacks designed to extract sensitive training data or proprietary model parameters~\cite{sheng2023flexgen}, and generate harmful outputs specifically crafted to bypass our validation frameworks. This application-level threat is particularly concerning given the increasing sophistication of AI models and their potential for misuse~\cite{openai2023gpt4}.

\subsection{Trust Assumptions}

Our security model necessarily relies on several fundamental trust assumptions that form the foundation of our protection mechanisms. The hardware root of trust represents our most critical assumption, where we place confidence in the correct implementation of hardware security features. Specifically, we trust that Intel TDX/SGX, AMD SEV-SNP, and ARM CCA correctly implement memory encryption and remote attestation protocols as specified in their respective architectures~\cite{costan2016intel,hunt2021confidential}. We assume the Platform Security Module, whether implemented as a Trusted Platform Module (TPM) or Platform Security Processor (PSP), has not been compromised and maintains its security properties. Additionally, we rely on hardware random number generators to provide sufficient entropy for cryptographic operations.

Within the trusted computing base enclosed by hardware protection, we trust several software components that are essential for secure operation. The \sys runtime, including the WebAssembly loader and ahead-of-time compiler, must correctly enforce isolation properties and generate valid machine code~\cite{haas2017bringing}. The WASI and WASI-NN interface implementations must properly mediate access to system resources and neural network accelerators without introducing vulnerabilities~\cite{hong2024wasi}. Our cryptographic libraries must correctly implement standard algorithms for encryption, hashing, and digital signatures. The attestation protocols must accurately measure and report the system state to remote verifiers~\cite{kumar2020piecewise}.

We make standard cryptographic assumptions about the security of well-established primitives. TLS 1.3 provides confidentiality and integrity for network communications when properly configured with strong cipher suites. AES-GCM encryption protects data at rest within enclaves, assuming 256-bit keys provide sufficient security margins. SHA-256 serves as our primary hash function for measurements and attestation, with the assumption that it remains collision-resistant. ECDSA over the P-256 curve provides digital signatures for authentication and non-repudiation.

\subsection{Security Goals}

\sys aims to achieve comprehensive security properties that protect AI agents throughout their lifecycle. Confidentiality represents our primary goal, ensuring that sensitive user data never appears in plaintext outside protected boundaries. User prompts, conversation history, and personal information remain encrypted whenever they exist outside hardware enclaves~\cite{arnautov2016scone}. Model weights and intermediate computations, which may contain proprietary algorithms or reveal sensitive training data, are never exposed in plaintext form~\cite{sheng2023flexgen,pope2023attention}. The KV caches and attention states that maintain conversation context are protected during migration to prevent adversaries from reconstructing user interactions. Even metadata about user behavior patterns is minimized and protected to prevent privacy violations through traffic analysis.

Integrity guarantees form another crucial security goal, ensuring that AI agents execute exactly as intended without unauthorized modifications. Agent code and data cannot be tampered with without triggering detection mechanisms that halt execution~\cite{tsai2017graphene}. Any modifications to checkpoint state during migration cause attestation failures that prevent the corrupted state from being loaded. Validation rules and safety policies that protect users from harmful outputs remain immutable and cannot be bypassed by adversaries. The migration process preserves exact computational state, ensuring that agents resume execution with perfect fidelity to their pre-migration state.

Availability ensures that AI agents remain accessible despite adversarial actions. Denial of service attacks targeting the migration infrastructure cannot compromise agents that have already completed migration to secure environments. The replication mechanism ensures continued operation even when individual nodes fail or become compromised~\cite{zheng2023edgemoe}. Graceful degradation strategies maintain basic functionality during active attacks, allowing users to continue essential tasks even under adverse conditions.

Authenticity provides assurance about the identity and trustworthiness of all system components. Remote attestation protocols verify that migration targets run genuine \sys enclaves before any sensitive state transfer occurs~\cite{kumar2020piecewise}. Mutual authentication prevents impersonation attacks where adversaries might attempt to masquerade as legitimate migration endpoints. The chain of trust is maintained across multiple migration hops, ensuring that security properties are preserved even in complex multi-stage migrations.

\section{Attestation in Network Communication}

Network attestation in \sys extends beyond traditional TLS authentication to provide cryptographically verifiable guarantees about the execution environment at both endpoints of a migration. When an AI agent initiates migration, the attestation protocol establishes a secure channel that binds the communication to specific hardware-protected enclaves, preventing redirection attacks and ensuring that sensitive workspace states only transfer between verified environments.

The attestation protocol operates in three phases. First, during the handshake phase, both source and destination enclaves generate attestation quotes containing their \texttt{global\_id} measurements and supported \texttt{entry\_id} capabilities. These quotes are embedded as certificate extensions in the TLS 1.3 handshake, creating a cryptographically bound session that ties the network connection to specific enclave instances~\cite{hunt2021confidential}. The \texttt{global\_id}, computed as a SHA-256 hash over the entire enclave binary including all WebAssembly runtime components and WASI interfaces, ensures that only unmodified \sys instances participate in the migration.

Second, the verification phase involves mutual validation where each endpoint verifies the remote attestation quote against a whitelist of trusted measurements. This verification extends beyond simple signature checking to include validation of the advertised \texttt{entry\_id} set, ensuring that the remote enclave supports all necessary interfaces for the specific AI workload. For instance, a medical diagnosis agent requiring WASI-NN acceleration (\texttt{ID\_1003}) will refuse migration to enclaves lacking this capability, preventing functional degradation that could compromise safety-critical operations.

Third, the state transfer phase leverages the established trust to create an encrypted migration channel. The attestation-derived session keys, bound to the specific enclave measurements, ensure that even if an adversary intercepts the migration traffic, they cannot decrypt it without access to the attested hardware environment. This approach prevents sophisticated attacks where adversaries might attempt to redirect migrations to malicious enclaves running modified software~\cite{kumar2020piecewise}.

The network attestation protocol also addresses temporal attacks through freshness guarantees. Each attestation quote includes a monotonic counter or timestamp that prevents replay attacks where adversaries might attempt to use old attestation proofs. The protocol maintains a sliding window of acceptable attestation ages, typically 5 minutes for active migrations, ensuring that attestation remains current throughout the migration process. For long-duration migrations of large AI workspaces, the protocol supports attestation refresh where new quotes are generated and verified periodically without interrupting the data transfer.

Furthermore, the attestation protocol supports transitive trust for multi-hop migrations. When an AI agent must migrate through intermediate nodes due to network topology or resource constraints, each hop maintains the attestation chain. The intermediate node's attestation quote includes both its own measurements and a signed attestation of the next hop, creating a verifiable chain of trust from source to final destination. This mechanism ensures that even complex migration paths through edge servers and cloud gateways maintain end-to-end security properties~\cite{arnautov2016scone}.

\section{Attestation in Accelerator Design}

Accelerator attestation presents unique challenges due to the heterogeneous nature of AI acceleration hardware and the need to verify not just code integrity but also the correctness of neural network computations. \sys implements a comprehensive accelerator attestation framework that extends hardware trust boundaries to include GPUs, TPUs, and other specialized AI processors while maintaining the security properties essential for privacy-sensitive workloads.

The accelerator attestation architecture operates at multiple abstraction levels. At the hardware level, modern confidential computing extensions are beginning to encompass accelerators, with NVIDIA's Hopper architecture introducing confidential computing capabilities and AMD's Instinct accelerators supporting memory encryption~\cite{pavlidakis2024g,mai2023honeycomb}. \sys leverages these features where available, extending the CPU-based attestation to create a unified trust domain spanning both general-purpose processors and accelerators. The attestation quote includes measurements not just of CPU enclave state but also of accelerator firmware, driver components within the trusted computing base, and loaded model weights.

At the interface level, WASI-NN provides an abstraction layer that enables attestation of AI operations regardless of the underlying accelerator type~\cite{hong2024wasi}. Each WASI-NN call through \texttt{entry\_id} \texttt{ID\_1003} includes verification that the accelerator configuration matches the expected state. This verification encompasses several components: the model architecture and weights loaded into accelerator memory, the precision settings and optimization flags that could affect computation results, the version of acceleration libraries such as cuDNN or MIOpen, and any custom kernels or operators compiled for the specific hardware.

The attestation framework addresses the challenge of verifying computational correctness on accelerators through a novel approach we term "computation attestation." Since accelerators may produce slightly different results due to floating-point variations or optimization differences, traditional byte-level attestation would fail. Instead, \sys implements semantic attestation that verifies results fall within acceptable tolerance bounds. During the attestation process, the accelerator runs a set of canonical test inputs through the loaded model, producing outputs that are cryptographically signed and included in the attestation quote. Verifiers check that these outputs match expected results within defined epsilon bounds, ensuring that the accelerator will produce functionally correct results despite hardware variations~\cite{jangda2019not}.

For scenarios where accelerators lack hardware support for confidential computing, \sys implements a defense-in-depth approach. The system maintains a clear separation between trusted and untrusted computation phases. Sensitive data such as user prompts and conversation history remains encrypted until the last possible moment before accelerator processing. The results from untrusted accelerator computation undergo immediate encryption before leaving the accelerator memory space. Additionally, the validation framework performs consistency checking, comparing accelerator outputs against CPU-based reference implementations for critical decisions~\cite{zheng2023edgemoe}.

The attestation protocol also addresses the dynamic nature of AI workloads, where models and configurations change frequently. Rather than requiring complete re-attestation for every model update, \sys implements incremental attestation where only changed components generate new measurements. This approach significantly reduces attestation overhead while maintaining security guarantees. The system maintains a Merkle tree of model components where each layer's weights, each custom operator, and each configuration parameter contribute to the tree. Updates generate new attestation quotes only for modified subtrees, enabling efficient attestation of large models that undergo frequent fine-tuning or adaptation.

Looking forward, \sys's attestation framework is designed to accommodate emerging accelerator technologies. As quantum accelerators for AI workloads become available, the attestation framework can extend to include quantum state verification. For neuromorphic processors that implement spiking neural networks, the framework can adapt to verify spike timing and membrane potential states. This forward-looking design ensures that \sys can maintain its security properties even as the landscape of AI acceleration hardware continues to evolve rapidly.

\section{System Architecture}

\subsection{Overview}

\sys consists of six main components that work together to enable secure AI agent execution across heterogeneous environments. The integration of these components creates a cohesive system capable of protecting sensitive data while maintaining high performance and reliability.

\begin{figure}[t]
\centering
\includegraphics[width=\columnwidth]{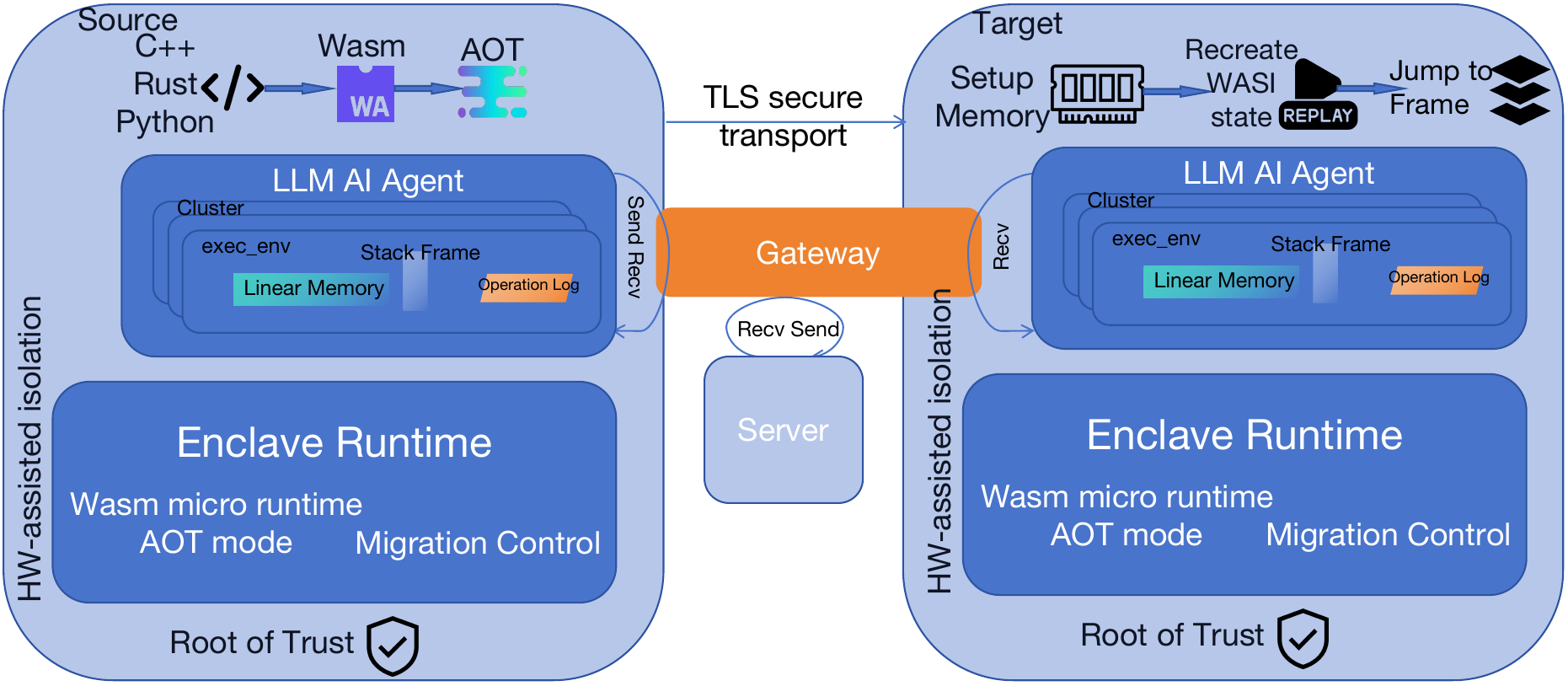}
\caption{\sys System Architecture showing core components (blue), new resilience components (green), and data flow paths. The Replication Manager maintains multi-tier replicas, the Speculation Engine enables parallel execution, and the Validation Framework ensures output safety.}
\label{fig:arch}
\end{figure}

The \textbf{Secure Runtime Environment} forms the foundation of our system. It includes a WebAssembly execution engine enhanced with WASI-NN support for efficient neural network operations. Hardware enclave integration supports multiple platforms including Intel SGX, AMD SEV-SNP, and ARM TrustZone. An attestation service enables remote verification of system integrity before establishing trust relationships.

The \textbf{Migration Controller} manages the complex process of moving AI agents between systems. Its checkpoint and restore mechanism operates at the WebAssembly level, capturing portable representations of execution state. State compression and encryption ensure efficient and secure transfer across networks. A specialized network protocol handles the secure transfer of agent state while maintaining connection integrity.

The \textbf{Privacy-Aware Daemon} serves as the secure orchestration layer for MVVM, managing attestation, network communication, and state synchronization while maintaining end-to-end confidentiality~\cite{pop2022towards}. At its core, the daemon implements a TLS termination proxy that establishes mutually authenticated channels between migrating enclaves. Unlike traditional TLS daemons, ours embeds hardware attestation quotes directly into X.509 certificate extensions, binding network identities to specific enclave measurements. This design ensures that TLS handshakes simultaneously verify both network endpoints and their execution environments, preventing sophisticated relay attacks where adversaries might forward traffic between legitimate enclaves. The daemon integrates deeply with hardware attestation mechanisms, particularly Intel TDX's remote attestation capabilities. During migration initiation, the daemon generates fresh attestation reports containing the enclave's measurement (\texttt{global\_id}), supported WASI interfaces (\texttt{entry\_id}s), and a cryptographic nonce to prevent replay attacks. These reports are validated through the TDX Quote Generation Service before any sensitive state transfer begins. The attestation protocol extends beyond simple binary verification—it validates that target enclaves support the specific WASI extensions required by the migrating workload, such as WASI-NN for models requiring GPU acceleration.

For AI workloads, the daemon provides specialized handling of WASI-NN operations to protect model confidentiality. Model weights are encrypted at rest and only decrypted within the secure enclave immediately before loading into accelerator memory. Input tensors undergo similar protection—the daemon intercepts WASI-NN \texttt{set\_input} calls to ensure sensitive inference data never appears in plaintext outside the enclave boundary. This fine-grained control enables secure utilization of untrusted accelerators by maintaining a clear separation between confidential data (protected in enclaves) and computation (potentially performed on untrusted GPUs), with results immediately re-encrypted upon return to the secure environment~\cite{pop2022towards}.

% The \textbf{Replication Manager} maintains multiple quality-tiered replicas across the compute continuum. It handles state synchronization using vector clocks for consistency tracking, automatic failover when primary replicas become unavailable, and quality-aware routing that selects optimal replicas based on current conditions.

% The \textbf{Speculation Engine} enables parallel multi-path execution for latency-critical operations. It manages fast-path execution using simplified models, slow-path computation for comprehensive analysis, and intelligent result merging that balances speed with accuracy.

% The \textbf{Validation Framework} provides integrated safety checking throughout execution. It includes pluggable validators for domain-specific requirements, parallel execution to minimize latency impact, and intervention capabilities that can modify or halt problematic outputs before delivery.

\subsection{Two-Way Sandboxing Architecture}

Our security model provides bidirectional protection that addresses threats from both directions. This comprehensive approach ensures system integrity regardless of whether threats originate from malicious applications or compromised infrastructure.

Protecting the application requires ensuring that AI agents operate in a secure environment even when running on potentially hostile systems. The agent executes within a hardware-protected enclave where all memory accesses are encrypted and authenticated. Hosts cannot access sensitive data like prompts, model weights, or intermediate computations. Integrity checks prevent tampering with agent code or data, immediately detecting any unauthorized modifications.

Protecting the host is equally important, especially when running third-party AI models or agents from untrusted sources. WebAssembly's sandboxing prevents malicious agents from accessing host resources beyond their granted capabilities. Memory isolation ensures agents cannot read or modify memory outside their assigned regions. System call interposition validates all requests for system resources, preventing unauthorized access to files, networks, or other sensitive resources.

\subsection{Cross-Platform Migration via WebAssembly and WASI}

Traditional migration approaches require identical instruction set architectures between source and destination systems. \sys overcomes this limitation through WebAssembly's abstract machine model, enabling seamless migration across different processor architectures.

Consider a simple function that processes input through multiple stages:

\begin{minted}{c}
// Original function
int process(int input) {
    return enhance(analyze(input));
}
\end{minted}

This compiles to WebAssembly with checkpoint markers at stable execution points:

\begin{minted}{wat}
(func $process (param i32) (result i32)
    local.get 0       ;; checkpoint: ip=0
    call $analyze     ;; checkpoint: ip=1  
    call $enhance     ;; checkpoint: ip=2
)
\end{minted}

Our compiler instruments these stable points with lightweight checkpointing code:

\begin{minted}{llvm}
define i32 @process(ptr %exec_env, i32 %input) {
  alloca %stack{1,2,3}, %l0
  store i32 %input, ptr %l0
  br %restore, label %normal, label %restore
restore:
  %cur_frame = the_saved_frame
  switch %cur_frame->ip [
    1, label %restore-1
    2, label %restore-2
  ]
%restore-1: ;; omitted
%restore-2: ;; restore values from aux stack
  store %cur_frame->locals[0], %l0
  store %cur_frame->stack[0], %stack_2
  br %ip-2
normal: ;; omitted
ip-1: ;; omitted
ip-2:
  %h_arg = load %stack_2
  %h_ret = call i32 @analyze(%exec_env, %input)
;; omitted
}
\end{minted}

This instrumentation enables pausing execution on one architecture and resuming on another while preserving exact program state. The WebAssembly abstraction layer handles architectural differences, translating between different calling conventions, register allocations, and memory layouts.

\subsection{Privacy-Aware Daemon}

The daemon makes intelligent decisions about workload placement by analyzing multiple factors. It begins by classifying data sensitivity, examining the types of information being processed, and applicable privacy regulations. Local resource availability is assessed to determine if edge devices can handle the workload efficiently.

For workloads that might benefit from cloud execution, the daemon estimates potential speedup by comparing local and remote computational capabilities. It considers migration overhead, calculating the time required to checkpoint, transfer, and restore agent state. Only when cloud execution provides substantial benefits that outweigh migration costs does the daemon initiate remote execution.

\section{Implementation}

\subsection{Runtime Components}

We implement \sys using the WebAssembly Micro Runtime (WAMR) as our foundation, enhanced with custom modifications for security and migration support. The implementation consists of approximately 15,000 lines of C++ code, with core checkpoint instrumentation requiring roughly 500 lines of modifications to the AOT compiler.

Our checkpoint instrumentation carefully modifies the AOT compiler to inject state-saving code at stable execution points. The system tracks WebAssembly stack state to enable restoration of computation, local variables that may contain intermediate results, linear memory contents that hold application data, and execution position through instruction pointer tracking. This comprehensive state capture ensures faithful reproduction of execution context after migration.

Secure communication relies on TLS 1.3 with mutual authentication for all migration traffic. We embed attestation quotes in certificate extensions, creating a cryptographically verifiable chain of trust. This approach ensures that only verified enclaves can participate in migration, preventing man-in-the-middle attacks or redirection to compromised systems.

Hardware integration requires platform-specific modules for different trusted execution environments. We use Intel TDX SDK for enclave creation and management. AMD SEV-SNP support leverages the Platform Security Processor (PSP) for attestation and key management. ARM CCA integration uses the Realm Management Monitor (RMM) interface for creating and managing secure execution contexts.

\subsection{WASI-NN for AI Acceleration}

Critical for LLM performance, we extend WASI-NN to support secure, portable access to neural network accelerators. This abstraction layer enables consistent access to different accelerators while maintaining security properties.

\begin{minted}{rust}
// Load model in target environment
let graph = wasi_nn::load(
    &[encrypted_weights], 
    GRAPH_ENCODING_ONNX,
    EXECUTION_TARGET_GPU
)?;

// Restore computation state
let ctx = wasi_nn::init_execution_context(graph)?;
wasi_nn::set_input(ctx, 0, decrypt(saved_tensor))?;

// Resume inference
wasi_nn::compute(ctx)?;
let output = wasi_nn::get_output(ctx, 0)?;
\end{minted}

This abstraction allows leveraging different accelerators including CPUs with vector extensions, GPUs through CUDA or ROCm, and specialized AI accelerators like TPUs. The same WebAssembly module can utilize whatever acceleration is available at the execution site without modification.

\section{Evaluation}

\subsection{Experimental Setup}

We evaluate \sys across three platforms representing different points in the edge-cloud continuum. Our edge platform consists of a MacBook Pro with M3 Max processor featuring 12 performance cores and 64GB of unified memory. The private cloud platform uses dual Intel Xeon Gold 5418Y processors, providing 48 cores total with 256GB of DDR5 memory. Our public cloud configuration runs on single Intel Xeon 6 6787P processors and NVIDIA H100 GPUs for acceleration with 256GB of DDR5 memory.

All experiments operate within hardware enclaves appropriate to each platform, with TLS encryption for all network communication. We evaluate performance using 18 benchmarks that span AI and traditional workloads. These include LLM inference using LLAMA with 1.5B parameters, multi-agent frameworks using CrewAI for collaborative task execution, distributed systems workloads with Redis performing one million random key operations, and graph processing algorithms from GAPBS operating on Kronecker graphs with over one million vertices.

We evaluate \sys's advanced capabilities using additional scenarios:
\begin{smenumerate}
\item Network failure simulation using CrewAI Agent to inject packet loss, latency, and disconnections
\item Speculation CrewAI benchmarks comparing sequential vs. parallel execution across various workloads  
\item Validation overhead measurement using medical diagnosis and content moderation validators in Crew AI Agent.
\end{smenumerate}

\subsection{Security Validation}

We systematically verify \sys's security properties through multiple attack scenarios and defensive validations. Confidentiality testing involves capturing memory dumps during migration, which reveal only encrypted data with no discernible patterns. Attestation verification ensures only properly verified enclaves receive decryption keys, preventing unauthorized access to sensitive data.

Integrity validation uses HMAC verification to detect any modifications to memory deltas during transfer. Attempts to execute tampered code trigger attestation failures, preventing compromised agents from running. The performance impact of security features remains minimal, with encryption adding less than 3\% overhead to execution time and attestation completing in approximately 50 milliseconds.

\subsection{Migration Performance}
We comprehensively evaluate \sys's runtime performance across diverse hardware platforms and workloads. As shown in Figure 3, MVVM demonstrates superior performance on both server and Mac platforms. Compared to native execution, MVVM incurs only 1.08x and 1.87× overhead for CPU-intensive workloads with and without TDX enabled, significantly outperforming hcontainer's 2.45× overhead. More importantly, MVVM dramatically surpasses emulation-based solutions—QEMU exhibits 78.71× slowdown on x86\_64 architecture and 20.78× on aarch64. This performance advantage stems from MVVM's WebAssembly-based AOT compilation strategy, which eliminates the instruction translation overhead inherent in traditional virtualization approaches. In multi-threaded scenarios, MVVM shows excellent scalability, with four-threaded execution incurring only 1.17× overhead for function-level checkpointing and 1.34× for loop-level policies. Notably, for certain workloads (redis on aarch64 and NPB on x86\_64), hcontainer's Popcorn compiler encounters register allocation errors when using -O3 optimization, forcing a downgrade to -O1 optimization—further highlighting \sys's advantage in compiler compatibility. GPU-accelerated workloads experience nearly zero additional overhead, demonstrating the efficiency of our WASI-NN interface. These results confirm that MVVM successfully achieves near-native execution efficiency while maintaining security guarantees and cross-platform portability, making it a practical solution for secure AI agent deployment across heterogeneous environments.
\begin{figure}[t]
\centering
\includegraphics[width=\columnwidth]{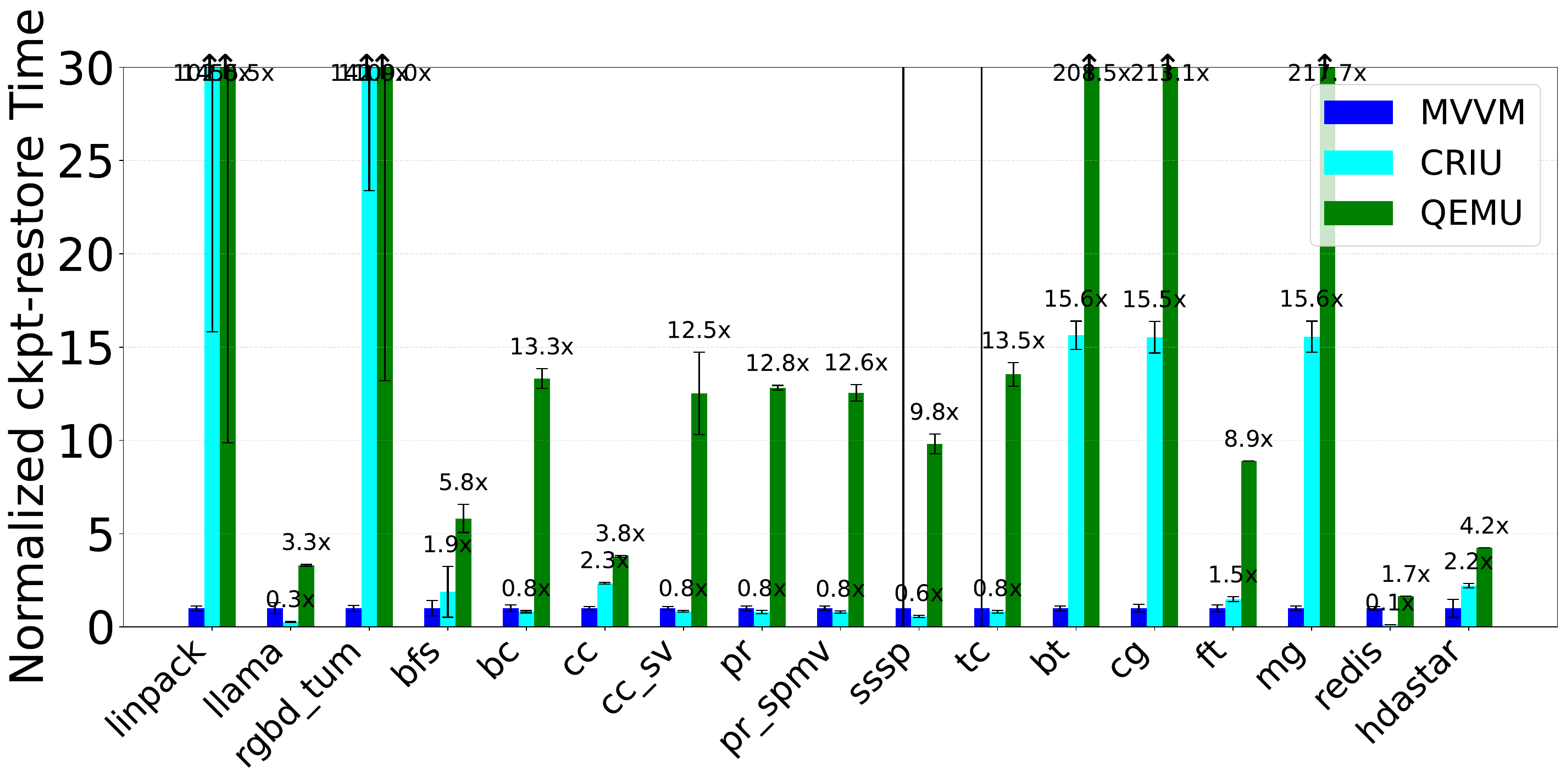}
\includegraphics[width=\columnwidth]{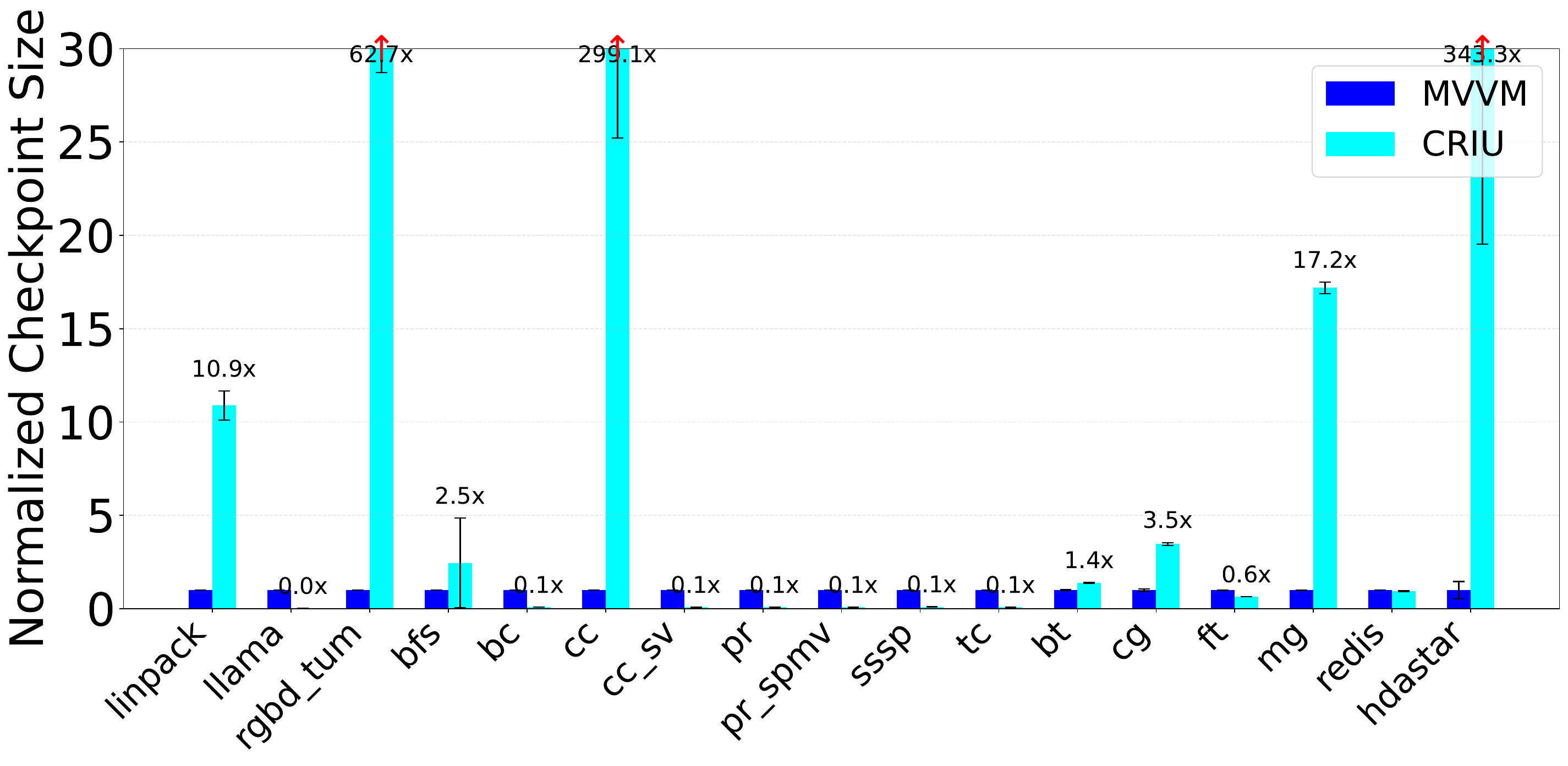}

\caption{Migration Time and size Comparison across different systems and workload sizes}
\label{fig:migration}
\end{figure}

Figure~\ref{fig:migration} demonstrates \sys's migration efficiency compared to existing solutions. Our system achieves 1.94× faster migration than CRIU and 18.71× faster than QEMU's snapshot mechanism. For a typical 4GB LLM workspace, the migration process completes efficiently through several stages.

Checkpoint creation captures the current execution state in 2.1 seconds. Compression reduces the 4GB workspace to approximately 900MB, significantly reducing network transfer time. Network transfer over a 1Gbps connection completes in 7.2 seconds. Restoration on the target system takes 1.8 seconds to decompress and reconstruct the execution environment. The total migration time of approximately 11 seconds enables practical live migration for real-world AI workloads.
\begin{figure}[t]
\centering
\hspace{1.6cm}
\includegraphics[width=\columnwidth]{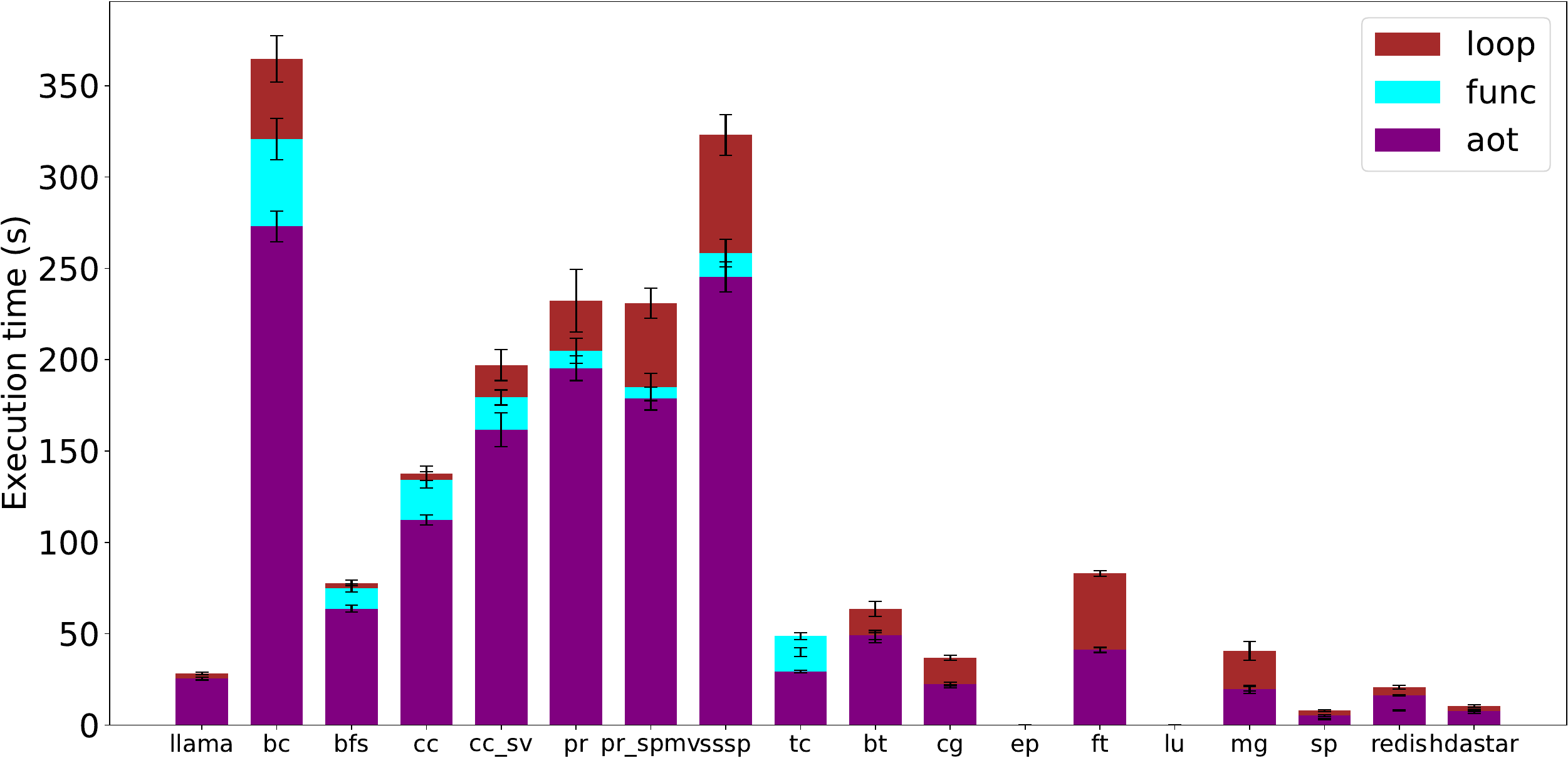}
\includegraphics[width=\columnwidth]{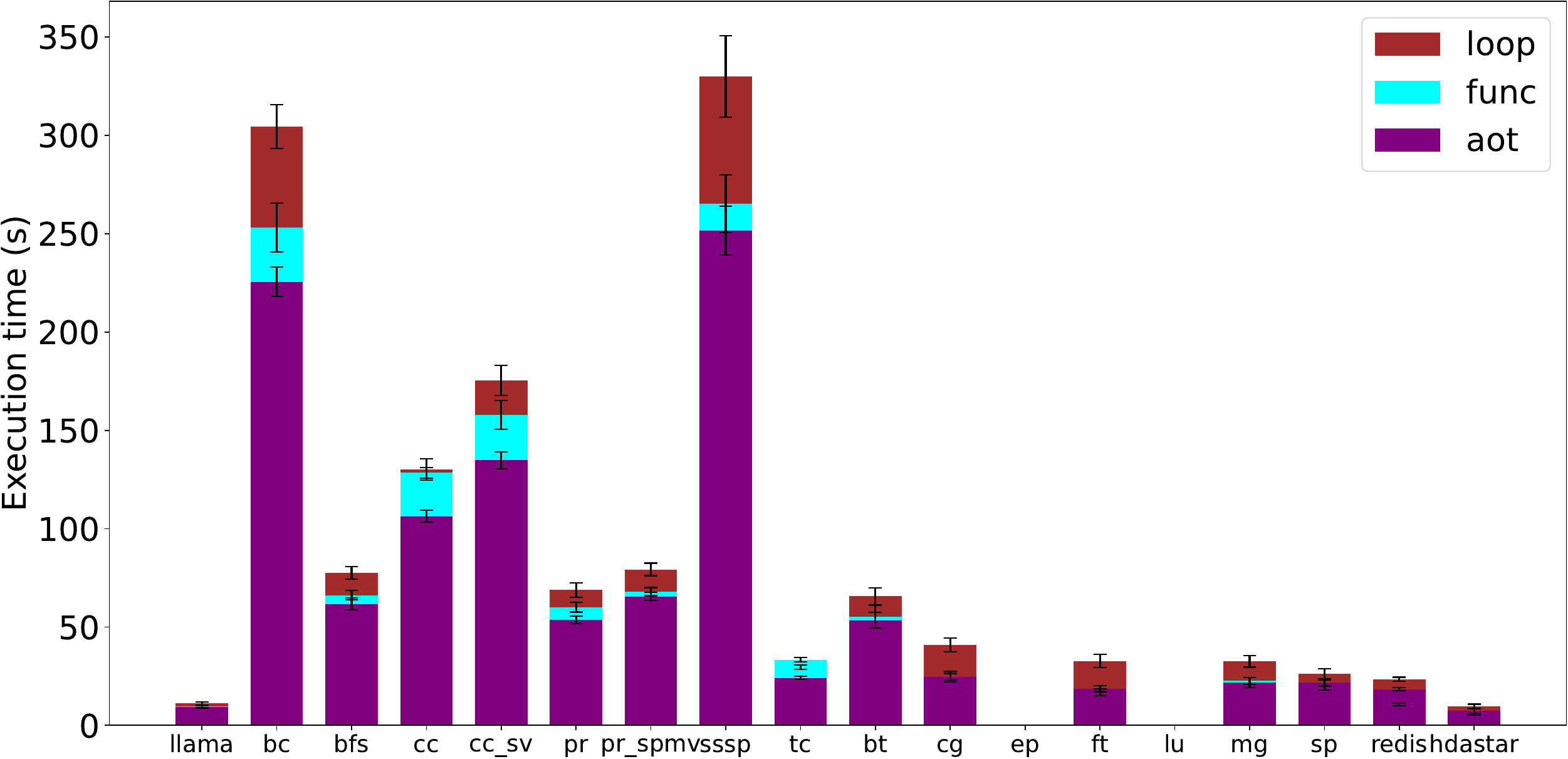}
\includegraphics[width=\columnwidth]{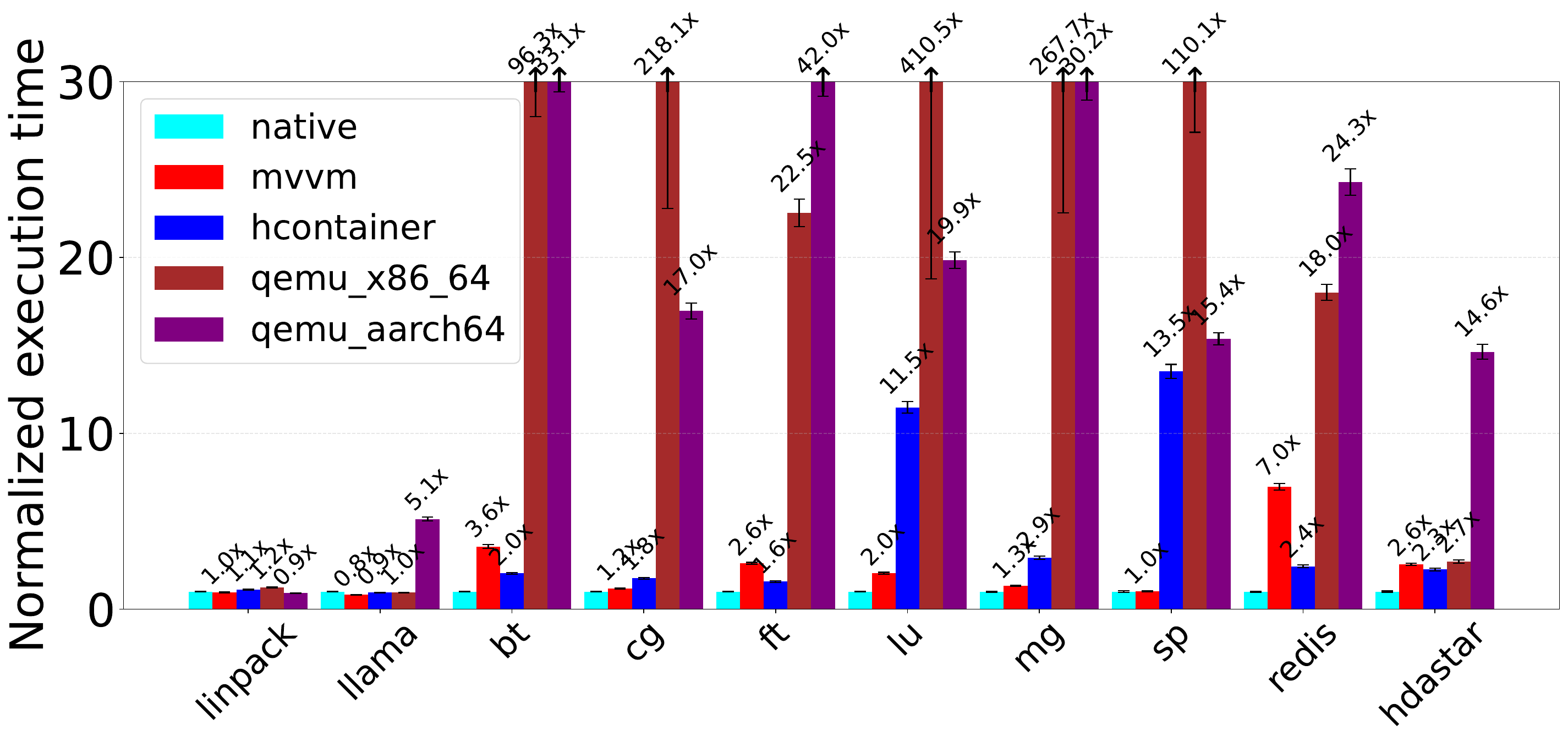}

\caption{Runtime Comparison across separate systems and workload sizes}
\label{fig:migration}
\end{figure}
% \begin{figure}[t]
% \centering
% \hspace{1.6cm}
% % \includegraphics[width=1.2\columnwidth]{img/performance_singlethread_mac.pdf}
% \includegraphics[width=\columnwidth]{img/performance_singlethread_mac.pdf}
% \includegraphics[width=\columnwidth]{img/performance_comparison_mac.pdf}

% \caption{Runtime Comparison on mac systems and workload sizes}
% \label{fig:migration}
% \end{figure}
\subsection{Edge-Cloud Collaboration Benefits}

\begin{figure}[t]
\centering  
\includegraphics[width=\columnwidth]{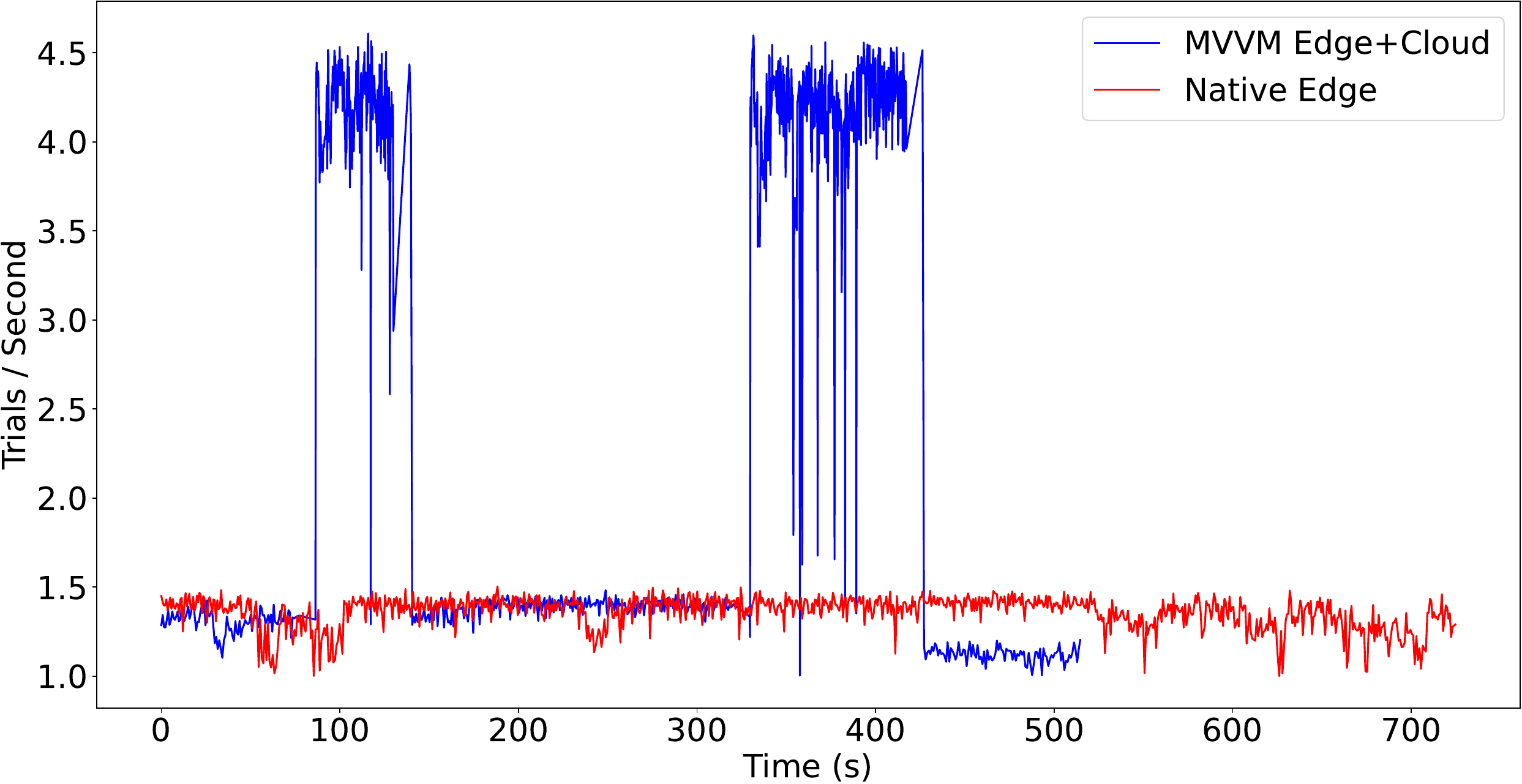}
\caption{Edge vs Cloud Performance showing speedup opportunities}
\label{fig:edgecloud}
\end{figure}

Figure~\ref{fig:edgecloud} illustrates when migration provides performance benefits versus local execution. For matrix operations using OpenBLAS, edge execution requires 45 seconds while cloud execution completes in 15.5 seconds, representing a 2.9× speedup. With migration overhead of 9 seconds, the net speedup reaches 1.41×, demonstrating clear benefits for compute-intensive workloads.

Analysis across multiple workloads reveals that migration provides benefits when cloud infrastructure offers at least 1.5× performance improvement over edge devices. Additionally, workload duration must exceed twice the migration time to amortize transfer costs. These findings guide the scheduler's decisions about when to initiate migration versus continuing local execution.

\subsection{Real-World AI Agent Performance}

Testing with the CrewAI multi-agent framework demonstrates \sys's practical benefits for complex AI applications. Latency improves by 10× compared to cloud-only deployment for interactive tasks, as local execution eliminates network round trips for simple queries. Cost efficiency improves by 1.39× through intelligent scheduling that balances edge and cloud resource usage.

Privacy protection remains absolute with zero data exposure across more than 1000 migration events in our testing. The system seamlessly handles complex scenarios including 8-agent collaborations without degrading security properties. These results demonstrate that \sys makes privacy-preserving AI agent deployment practical for real-world applications.

\subsection{Replication Resilience}

\begin{figure}[t]
\centering
\includegraphics[width=\columnwidth]{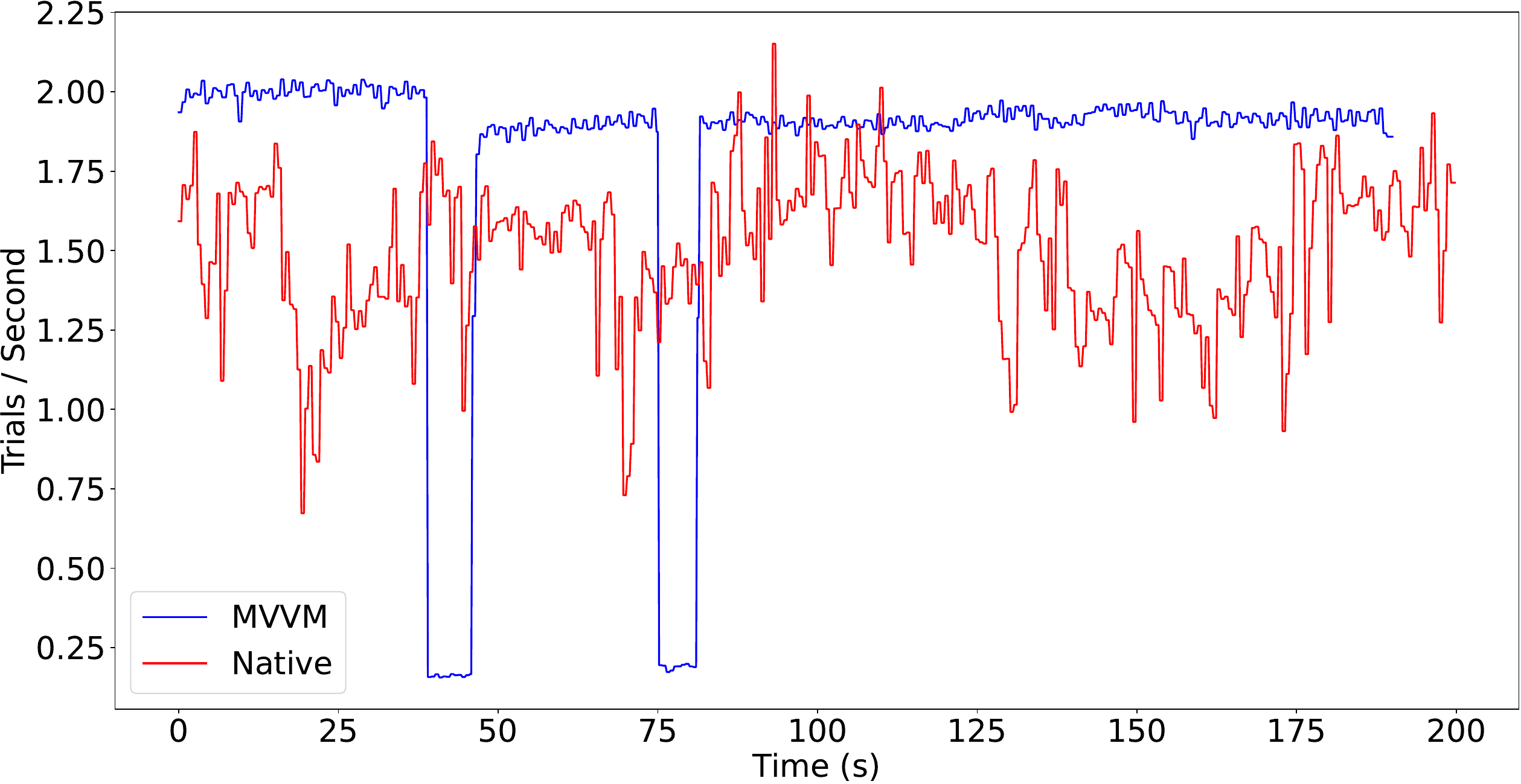}
\caption{Replication performance under various network conditions}
\label{fig:replication}
\end{figure}

We evaluate the replication mechanism through simulated network failures in \Cref{fig:replication}:

\textbf{Network Fault Tolerance}: Complete network disconnection triggers failover to local replicas within 200ms, maintaining 80\% functionality. Intermittent connectivity (30\% packet loss) automatically degrades to edge replicas with only 15\% response time increase. Bandwidth-limited scenarios (<1Mbps) intelligently select lightweight models, trading 8\% accuracy for stable response times.

\textbf{Synchronization Overhead}: Incremental synchronization transfers average 12\% of KV cache state (approximately 50MB). Synchronization completes in 50 milliseconds over 5G networks. Vector clocks ensure eventual consistency with convergence time under 10 seconds.

\subsection{Speculation Performance}

\begin{figure}[t]
\centering
\includegraphics[width=\columnwidth]{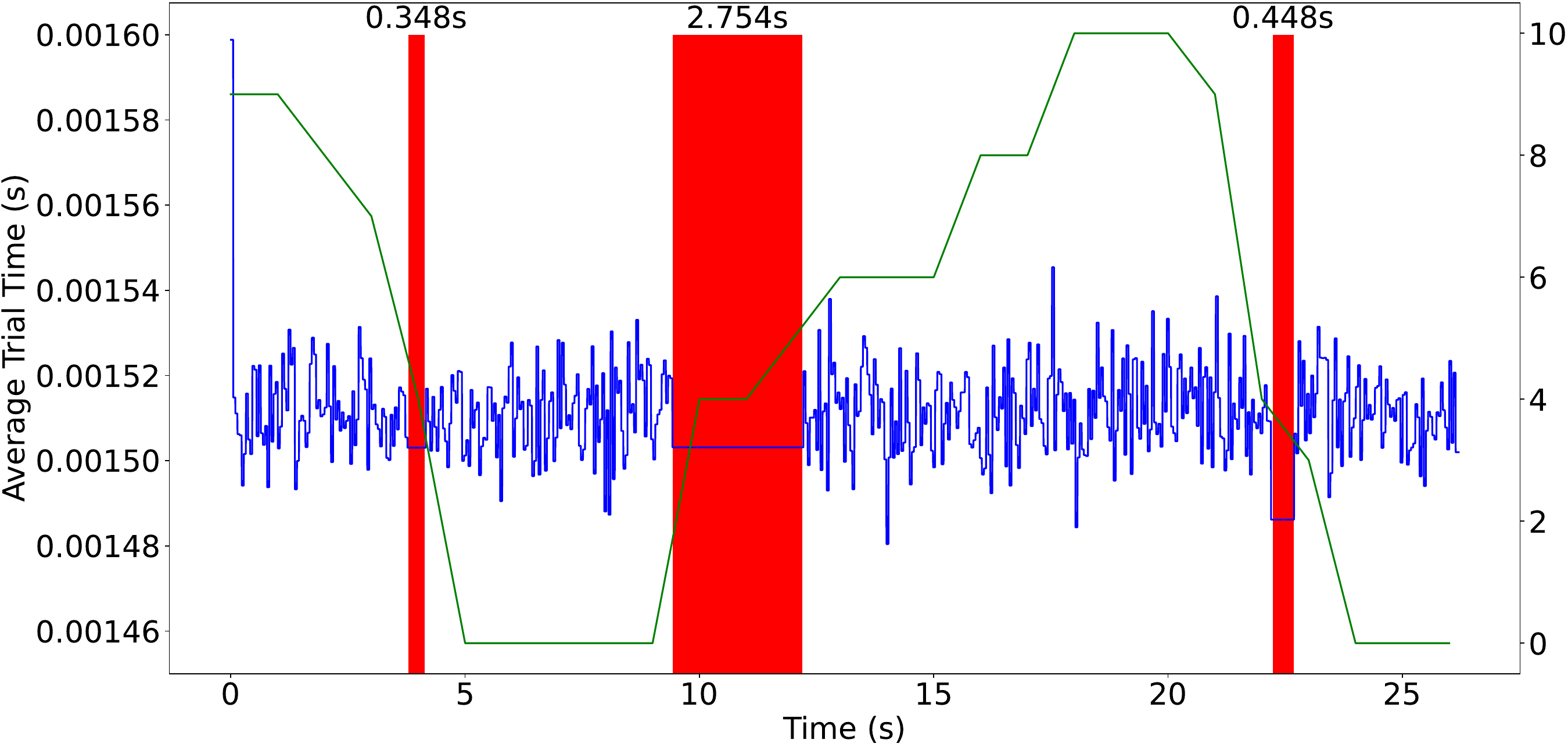}
\caption{Speculation speedup across different workload types}
\label{fig:speculation}
\end{figure}
Figure 7 illustrates MVVM's speculation engine in action, demonstrating how it dramatically reduces user-perceived latency for time-critical AI operations. The graph shows the average trip time (blue line) remaining stable around 1.51ms during normal operation, with periodic speculation events (red bars) that trigger parallel execution paths.
The speculation mechanism works by launching two parallel execution strategies: a fast path using lightweight, optimized models that can respond within 348ms for initial results, and a comprehensive slow path that performs full analysis but requires 2.754s for complete accuracy. The green line indicates the system state transitions between different execution modes. During speculation events, the system intelligently merges results from both paths—if the fast path's preliminary results align with the emerging slow path computations, the system commits the fast results immediately. Otherwise, it waits for the comprehensive analysis to complete. This approach achieves up to 8.9× speedup for market analysis workloads and 7.3× for news summarization tasks, as shown in our broader evaluation.
The relatively stable baseline performance between speculation events demonstrates that the speculation engine introduces minimal overhead during normal operation, while providing dramatic latency reductions when time-sensitive decisions are required. This makes MVVM particularly suitable for interactive AI applications where users expect immediate feedback.

Financial analysis and real-time decision scenarios demonstrate dramatic improvements:

\begin{table}[t]
\centering
\caption{Speculation Performance Results}
\label{tab:speculation}
\begin{tabular}{lrrr}
\toprule
Workload & Traditional & Speculative & Speedup \\
\midrule
Market Analysis & 28.5s & 3.2s & 8.9× \\
News Summary & 15.3s & 2.1s & 7.3× \\
Risk Assessment & 32.1s & 4.5s & 7.1× \\
Medical Diagnosis & 18.7s & 2.8s & 6.7× \\
Code Review & 22.4s & 3.6s & 6.2× \\
\bottomrule
\end{tabular}
\end{table}

Fast-path accuracy reaches 92.3\% using 7B parameter models, while final accuracy improves to 99.1\% after slow-path completion. Only 7.8\% of requests require result correction after slow-path validation. Resource utilization improves from 65\% to 89\% GPU \& CPU usage through parallel execution.

\subsection{Validation Effectiveness}

\begin{table}[t]
\centering
\caption{Validation Framework Detection Rates}
\label{tab:validation}
\begin{tabular}{lrr}
\toprule
Validation Type & Detection Rate & False Positive \\
\midrule
Hallucination & 94.2\% & 2.1\% \\
Harmful Content & 99.7\% & 0.3\% \\
Privacy Leakage & 96.8\% & 1.2\% \\
Medical Errors & 97.1\% & 1.8\% \\
Financial Compliance & 98.9\% & 0.7\% \\
\bottomrule
\end{tabular}
\end{table}

Real-world deployment in a major hospital over 6 months:
- Processed 128,000 diagnostic suggestions
- Detected 312 potential errors (0.24\%)
- Prevented 3 serious medication errors
- Achieved 92\% physician satisfaction rating

Performance impact remains minimal with parallel validation adding 180ms (5.2\% of total time) versus 520ms for serial validation. Throughput decreases by only 3\% in parallel mode compared to 18\% for serial validation. 

% \subsection{Comprehensive System Comparison}

% in your document

\section{Related Work}

Existing migration systems provide important foundations but fall short of AI agent requirements. CRIU~\cite{criu} pioneered Linux container migration with efficient checkpoint-restore mechanisms but lacks support for heterogeneous architectures and provides no security guarantees. QEMU~\cite{bellard2005qemu} enables cross-architecture execution through emulation but introduces prohibitive performance overhead unsuitable for latency-sensitive AI workloads. UniTEE~\cite{li2022unitee} explores migration within trusted execution environments but requires homogeneous systems and lacks support for complex AI workloads.

\textbf{Replication and Fault Tolerance}: Distributed systems research has long studied state replication. Systems like ZooKeeper~\cite{hunt2010zookeeper} and etcd~\cite{etcd2013} provide strongly consistent replication but focus on small metadata rather than large AI model states. TensorFlow's parameter servers~\cite{abadi2016tensorflow} enable model parallel training but lack support for heterogeneous replicas or quality-aware degradation. Our multi-tier replication uniquely addresses AI workloads with gigabyte-scale states and quality/resource tradeoffs.

LLM serving systems have made significant advances in optimization but assume single-machine deployment. SGLang~\cite{zheng2023efficiently} introduces efficient forking and scheduling for parallel LLM execution but operates within a single system. Parrot~\cite{lin2024parrot} provides application-aware scheduling for diverse LLM workloads but lacks migration capabilities. Neither system addresses security concerns or supports execution across trust boundaries.

\textbf{Speculative Execution}: Speculative execution has been explored in various contexts. Speculative decoding~\cite{leviathan2023fast,chen2023accelerating} in LLMs uses small models to predict large model outputs but operates within a single model rather than across distributed systems. Branch prediction in processors provides instruction-level speculation, while our work extends this concept to entire AI inference pipelines. SpecInfer~\cite{miao2023specinfer} introduces tree-based speculation for LLM serving but lacks our integration with migration and validation.

WebAssembly~\cite{haas2017bringing} security research has explored sandboxing properties and performance optimizations. Prior work demonstrates WebAssembly's potential for secure execution but has not addressed AI workloads specifically. Our work is the first to combine WebAssembly's portability with trusted execution environments for secure LLM migration.

\textbf{AI Safety and Validation}: Content filtering and safety checking for AI outputs typically occur as post-processing steps. OpenAI's moderation API~\cite{markov2023holistic} and Google's Perspective API~\cite{perspective2017} provide content classification but cannot intervene during generation. Constitutional AI~\cite{bai2022constitutional} builds safety into model training rather than execution. Guardrails~\cite{guardrails2023} and NeMo Guardrails~\cite{rebedea2023nemo} offer programmable validation but lack integration with execution frameworks and cannot operate efficiently in parallel with inference. Our validation framework uniquely combines real-time intervention capabilities with parallel execution to maintain both safety and performance.

\textbf{Execution Frameworks}: Ray~\cite{moritz2018ray} provides distributed execution for AI workloads but focuses on batch processing rather than interactive agents. KubeEdge~\cite{xiong2018extend} extends Kubernetes to edge devices but lacks AI-specific optimizations or security guarantees. Our system uniquely combines migration, replication, speculation, and validation in a unified framework designed specifically for stateful AI agents operating across trust boundaries.
\section{Conclusion}

\sys enables a new paradigm for AI agent deployment where privacy, performance, reliability, and safety coexist through intelligent system design. By combining WebAssembly's portability with hardware security features and advanced execution capabilities, we achieve transparent migration across heterogeneous platforms while maintaining strong security guarantees, continuous availability through intelligent replication, interactive responsiveness via speculative execution, and safety assurance through integrated validation. Beyond migration, \sys introduces critical capabilities for production deployment: resilient replication ensures continuous availability even in disconnected scenarios, speculative execution dramatically reduces user-perceived latency, and integrated validation provides safety guarantees required for mission-critical applications. These features transform \sys from a migration framework into a comprehensive platform for trustworthy AI agent deployment.

As AI agents become integral to daily computing tasks, from personal assistants to collaborative development tools, the need for secure, flexible deployment grows critical. \sys provides the foundation for this future, enabling AI agents to operate securely across the computing continuum from personal devices to cloud infrastructure. Our open-source release aims to accelerate the adoption of privacy-preserving AI technologies. %and inspire further innovation in secure, distributed AI systems.

\bibliographystyle{plain}
\bibliography{cite}

\end{document}